\documentstyle[11pt,newpasp,twoside]{article}
\def\edcomment#1{\iffalse\marginpar{\raggedright\sl#1\/}\else\relax\fi}
\reversemarginpar

\begin{document}
\title{Discussion:  Star Formation Within Galaxies}
 \author{M. S. Oey}
\affil{Lowell Observatory, 1400 W. Mars Hill Rd., Flagstaff, AZ\ \ \
 86001, USA}
\author{C. Mu\~noz-Tu\~n\'on}
\affil{Instituto de Astrof\'\i sica de Canarias, C/ V\'\i a L\'actea s/n,
E-38200 La Laguna -- Tenerife, Spain}

\begin{abstract}

This Discussion session focused on star formation within galactic
scales.  We attempt to identify the dominant physical processes and
parameters that characterize star formation, and to identify key
questions that illuminate these phenomena.  The Discussion was
delineated by the following cycle of three questions:  (A) Is the top
of the {\sc H\thinspace ii} LF physically distinct?\ \  (B)  How does massive star
feedback affect the ISM and star formation?\ \ (C) How do ISM
properties affect the {\sc H\thinspace ii} LF?  Finally, is one of these three
questions the fundamental one of the cycle?

Corresponding answers emerged from the Discussion:  (A) The {\sc H\thinspace ii} LF
to date is a continuous power law;  (B) There are both positive and
negative feedback effects, which are poorly understood;  (C) The {\sc H\thinspace ii}
LF appears remarkably independent of ISM properties.  Therefore, we
suggest that the resultant fundamental question is:  {\bf ``Is the {\sc H\thinspace ii}
LF and parent stellar cluster membership function universal?''}  This
is analogous to the related question of a universal stellar IMF.  
Understanding the relationship, if any, between the IMF,
cluster membership function, and ISM properties may finally lead to a
quantitative theory of star formation.

\end{abstract}

\section{Introduction}

As is suggested by the title of this conference, star formation is one
of the fundamental drivers of evolution in the Universe.  It can be
considered on scales ranging from individual stars, to starbursts, to
the cosmic history of star formation itself.  Roberto Terlevich has
spent much of his career considering this entire range of star formation
events and interpreting the underlying physical processes that drive
the evolution of galaxies and the Universe. 

Star forming events are telling us many things, some of which we understand,
and many of which we still do not.  In this sense, there are some key works
that attempt fundamental interpretations of information from star-forming  
regions.  Roberto has been a leader in this approach, and he,
together with Jorge Melnick, became a front runner when proposing the
``Size -- Velocity Dispersion'' diagramme applied to giant {\sc H\thinspace ii}
regions (G{\sc Hii}Rs).  The Terlevich \& Melnick law (Terlevich \& Melnick
1981) shows the existence of a 
correlation between radius and velocity dispersion in G{\sc Hii}Rs,
following a $ R \sim \sigma^{2}$ relation, the same as that for
globular clusters and elliptical galaxies.  Also the luminosity $L$ and
$\sigma$ are related following the relation, $L\sim \sigma^{4}$.
Despite the use of the diagrammes as distance indicators, as was their
initial attempt, their interpretation of the big
star-forming regions was important.  From their findings, Terlevich \&
Melnick proposed that G{\sc Hii}Rs are virialized systems like globular
clusters and elliptical galaxies, and the measured diagnostics indicate
the mass of the total system. 

Since then there has been much debate on the subject, both on the
existence of the correlations themselves and on the interpretation
provided.  To reproduce the correlation and to confirm its validity have
involved many of us for a number of years.  We have been exploring the
definition of the parameters, namely the radius, which is crucial to infer
the other two, as well as the shape of the emission lines, which are
used to get the $\sigma$ value (see Melnick et al. 1988;
Mu\~noz-Tu\~n\'on 1994; Mu\~noz-Tu\~n\'on et al. 1995 and
references therein). 

Later on, the Terlevich \& Melnick law was extended to {\sc H\thinspace ii} galaxies
(Melnick et al. 1988; Telles \& Terlevich 1993), and it was demonstrated that
under some restriction on surface brightness (Fuentes-Masip et al. 2000) and 
analysis of the line profiles, the correlations still hold (see
Telles et al. 2001; and Telles, this volume).
This may indicate that the process of massive cluster formation is similar
in both {\sc H\thinspace ii} galaxies and G{\sc Hii}Rs.

The interpretation of the supersonic broadening of emission lines (first
reported by Smith \& Weedman 1970) and its correlation with size is still
an issue of debate.  Stellar winds have been blamed for the supersonic
width (see, e.g., Chu \& Kennicutt 1994, debated in Tenorio-Tagle et
al. 1996).  Supersonic turbulence is also proposed 
as a responsible mechanism for the correlations and line broadening, and
much work has been carried out both observationally (see the pionering work
by Casta\~neda 1988; and more recently by Joncas 1999, Miesch et al.
1999, and references therein) and theoretically (see 
V\'azquez-Semadeni 1999) to address this issue, which remains
open.

What is interesting although a bit worrisome, is that the Terlevich \&
Melnick laws are used nowdays to infer physical parameters (namely the
mass) of objects at high redshift.  Somehow, and despite the
controversies mentioned above, they have been successful in breaking
out to become one of the important tools for interpreting
star-forming galaxies at high $z$. 

\section{The Discussion}

In this discussion session, we continue in the spirit of searching for
fundamental interpretations of star-forming regions.
Here, we consider star formation within galactic
scales.  We attempt to make sense of the various phenomena related to star
formation, and we try to identify the dominant processes that affect it.
In organizing the topics for the discussion, each of us (CMT \& MSO)
independently compiled a list of our favorite relevant questions.  We
found that these issues all appear to fall within the confines of the
following three major questions:
\pagebreak

{\bf
\begin{itemize}
\item A.  Is the top of the {\sc H\thinspace ii} LF ({\sc H\thinspace ii} LF) physically distinct?  What
determines spatial concentration of star formation and starbursts?
\item B.  How does massive star feedback affect the ISM and star
formation?
\item C.  How do ISM properties such as metallicity, clumpiness, turbulence,
and phase balance affect the {\sc H\thinspace ii} region luminosity function and star formation?
\end{itemize}
}

These three questions form a cycle:  The {\sc H\thinspace ii} LF and especially,
the most luminous star-forming regions, determine the nature of
massive star feedback; feedback drives properties of the interstellar
medium (ISM); and ISM properties presumably must affect the {\sc H\thinspace ii} LF.
Is one of these three questions, A, B, or C, the fundamental
determinant of the other two, and thereby of galactic star formation
in general?  How did this seemingly chicken-and-egg cycle originate?
Are there actually external relevant issues that break the cycle?
The discussion session raised a number of observational and
theoretical issues that address these questions.  In \S 6 below, we
nominate our choice for The Fundamental Question.

\section{A:\ \  The {\sc H\thinspace ii} LF and starbursts}

The {\sc H\thinspace ii} region luminosity function is remarkably robust.
Observations over a wide range of galaxy types consistently show that
the differential LF is described by a power law of slope $-2\pm 0.3$.
This is found for disk galaxies (e.g., Kennicutt, Edgar, \& Hodge 1989;
Banfi et al. 1993; Rozas et al. 1996), including those with active
nuclei (Gonz\'alez-Delgado \& P\'erez 1997).  Oey \& Clarke (1998) show
that variations in the {\sc H\thinspace ii} LF, including observed slope breaks and
the apparently steeper slopes of Sa galaxies (Caldwell et al. 1991)
are almost all consistently explained by a universal power law,
\begin{equation}
N(N_*)\ dN_* \propto N_*^{-2}\ dN_* \quad ,
\end{equation}
for the number of clusters having $N_*$ ionizing stars in the
range $N_*$ to $N_* + dN_*$.  The only apparent exception appears to
be somewhat flatter slopes of $-1$ to $-1.5$ found in a study of dwarf
irregular galaxies (Youngblood \& Hunter 1999), although the {\sc H\thinspace ii} region
statistics in these small galaxies are more difficult.  The universal
power-law $N_*^{-2}$ is also seen for stellar clusters themselves
(Elmegreen \& Efremov 1997).

But what about the upper limit to the {\sc H\thinspace ii} LF?  We see that it can
vary between different galaxies:  Oey \& Clarke (1998) and Kennicutt
et al. (1989) find that early-type spirals show a cut-off in the
maximum luminosities of the star-forming regions around H$\alpha$
luminosities of $\log L_{\rm H\alpha} \sim 38$ whereas late-type
galaxies do not appear to show any maximum $L_{\rm H\alpha}$, often
having {\sc H\thinspace ii} regions with $\log L_{\rm H\alpha}\sim 41 - 42$.  In the
most actively star-forming galaxies, are the most vigorous
star-forming regions a physically distinct class of objects from the
rest of the nebular population?

The most active star-forming regions in many disk galaxies are the
circumnuclear regions, which are often associated with bar
activity.  Although the central galactic areas should host conditions
that are significantly different from the remainder of the disk,
Alonso-Herrero \& Knapen (2001) found that the {\sc H\thinspace ii} LFs for
circumnuclear {\sc H\thinspace ii} regions in 52 galaxies nevertheless show remarkably
normal slopes.  Thus, these luminous star-forming regions simply
represent an extension to the {\sc H\thinspace ii} LF of the disks.
Alonso-Herrero reports that preliminary analysis of photometric
broadband observations of luminous infrared
galaxies thus far confirms that the parent stellar
clusters likewise appear to be scaled-up versions of those found 
in less luminous {\sc H\thinspace ii} regions.  Direct observations of super star
clusters in starburst galaxies also confirms the $N_*^{-2}$ power law
in this regime (Meurer et al. 1995).

Nevertheless, it is apparent that the circumnuclear {\sc H\thinspace ii} regions are
systematically far brighter than the disk objects.  This implies that
nuclear conditions favor the formation of the most luminous regions.
Another environment that favors highly luminous star formation is
found in gas-rich dwarf galaxies, which also host some of the largest
star-forming regions.  The high-density, high-pressure, high-shear
environment near galactic nuclei would appear to greatly contrast with
the environment typically associated with the ISM in dwarf galaxies.
Could these two types of environment have common dominant variables
that induce energetic star formation?  Are there differences between
the giant {\sc H\thinspace ii} regions formed in these respective conditions? From
the discussion, it emerged that 
the formation of these luminous objects is probably determined by more
complex factors than a single determining parameter like gas density,
or else it would have been apparent by now.  More likely, a
combination of factors may dominate, for example gas density and
freefall timescale, which together could produce similar conditions
and high star formation in both of the environments identified above.

Another factor that must clearly limit the maximum {\sc H\thinspace ii} region
luminosities is the physical size of the host galaxy.  The largest
scales of star formation approach the physical scales of the host
galaxy, for example in starbursts and {\sc H\thinspace ii} galaxies.  Does this
factor affect the properties of the most luminous {\sc H\thinspace ii} regions?
There appears to be no evidence to date that it does.  This suggests
that star formation can be a highly localized effect, that need not be
strongly influenced by the extended ISM properties of the host
galaxies. 

Moreover, evolution clearly affects parameter measurements of
massive complexes.  {\sc H\thinspace ii} regions evolve, both in luminosity and shape.
In nearby and resolved systems we see nets of loops, shells, filaments
(for example, as seen in 30 Dor in the LMC or NGC 604 in
M33), which are undoubtedly probes of their advanced stage of evolution (see
Mu\~noz-Tu\~n\'on et al. 1996 for a discussion).  Therefore,
determination of the parameters relating to the most massive regions
should include observations at earlier evolutionary stages (e.g., IR
sources), in order to properly understand the upper limit of the
the {\sc H\thinspace ii} LF.  At present, there may exist an important bias towards
more evolved systems.

\section{B:\ \  Massive star feedback and the ISM}

The properties of massive star feedback vary, depending on the
characteristics of the parent star formation, and host galaxy ISM and
environment.  For ordinary star-forming galaxies, continuous low-level
or moderate star formation following the {\sc H\thinspace ii} LF may produce only
pockets of hot ($10^6$ K) gas in supernova-driven superbubbles, and
some spatially scattered nebulae within a diffuse warm ($10^4$ K)
ionized medium (WIM).  However, feedback from high star-formation rates can
be strongly dependent on, e.g., the spatial distribution of the star
formation, and the properties of the surrounding galactic halo or
intergalactic medium (IGM).  Clarke \& Oey (2002; see Oey \& Clarke,
this volume) show that, for the same high star formation rate, events
that are spatially distributed and following the {\sc H\thinspace ii} LF will shred the
neutral ISM into worms and filaments, thereby strongly enhancing the
escape of ionizing photons and metals from galaxies.  Centrally
concentrated star formation, on the other hand, would allow the escape
of photons and material only through the opening angle of
the central superwind cone.  The ISM in the latter case would remain
largely intact outside the central region.  Similarly, the fate of hot
gas and newly synthesized metals depends not only on the luminosity of
the star-forming regions relative to the galactic gravitational
potential, but also on the halo and external medium.  Silich \&
Tenorio-Tagle (2001) and Kunth et al. (2002) emphasize the difficulty
in ejecting material in the presence of external IGM pressure (see
also the Tenorio-Tagle \& V\'\i lchez Discussion session, this volume).

Whether or not photons and material escape from galaxies will
profoundly affect the ISM of the parent galaxies themselves.  If
ionizing photons cannot escape, then they are absorbed in the ISM and
contribute to a more strongly ionized diffuse WIM.  Presumably this
heating will act to inhibit star formation.  If hot gas and metals
cannot escape the galaxy, they too, are returned to the host ISM.  The
metals will enhance cooling and thereby presumably enhance star
formation; the hot gas will enhance mixing of these metals throughout
the galaxy, depending on its spatial distribution.  In short, the
phase balance of the ISM is largely determined by the character of the
massive star feedback.  Interstellar turbulence is another consequence
of mechanical feedback that must strongly affect star formation.
While turbulent motion itself inhibits gravitational collapse,
turbulence also restructures the multiphase ISM, perhaps promoting the
formation of cold clouds.

On global scales, the above arguments suggest both negative and
positive effects of feedback on star formation.  Which dominates?
On more localized scales, there are well-documented examples in the
Large Magellanic Cloud of sequential star formation, apparently
triggered by expanding superbubble shells (e.g., Dopita et al. 1985;
Parker et al. 1992; Oey \& Massey 1995).  Circumnuclear rings or ring
galaxies show evidence of propagating star formation, for
example, NGC 1068 (Myers \& Scoville 1987), probably triggered by
nuclear activity; and the Cartwheel Galaxy (Marcum et al. 1992),
triggered by a galaxy merger.  D\'\i az reports tentative evidence of
both radial and azimuthal age gradients in circumnuclear {\sc H\thinspace ii}
regions.  Yet, as often argued for star formation
at the molecular cloud scale, massive star feedback is also thought to
destroy clouds and suppress star formation.  On global scales,
feedback in dwarf galaxies may result in the blowaway of a significant
fraction of the ISM, also inhibiting star formation.  Understanding
how feedback affects star formation on both large and small scales
remains a key open question. 

\section{C:\ \ The ISM and the {\sc H\thinspace ii} LF}

How do ISM properties yield the $N_*^{-2}$ power law and resulting
{\sc H\thinspace ii} LF?  This is a question that parallels the long-standing problem
regarding the origin of the stellar initial mass function (IMF).
Presumably fundamental properties of the ISM determine the {\sc H\thinspace ii} LF.
However, the above discussion in \S 3 emphasized that $N_*^{-2}$ power
law is remarkably constant, even for extremely different populations of
objects, including low-luminosity {\sc H\thinspace ii} regions in early-type
galaxies, and high-luminosity circumnuclear regions.  This suggests
that the conditions causing the parent cluster membership function are
remarkably insensitive to ISM conditions.  Since fundamental ISM
parameters are density, pressure, and temperature, Lynden-Bell offers
the analogy of star formation as a phase transition.  A more complex
phase transition than that of simple matter, to be sure, but perhaps a
secure place to return to first principles?

Melnick noted the similarity between the $N_*^{-2}$ power law 
and the IMF power-law index of $-2.35$ (Salpeter 1955).  Fractal structure 
has long been a popular model for the ISM, and he suggested that the --2
power law simply results from a fractal default for complexity in
nature.  A difficult aspect of this model is that it appears to be
unrelated to any specific physics, and therefore can only be tied to
physical properties by circumstantial evidence.  One of the most popular
links is between fractals and interstellar turbulence (e.g., Norman \&
Ferrara 1996; see Oey 2002), since the latter yields a similar power
law for the spatial power spectrum.  However, it is now emerging that
much of the large-scale neutral hydrogen in the ISM is strongly
filamentary (Braun 1997; Elmegreen et al. 2001) which is difficult to
reconcile with a simple fractal structure (Elmegreen et al. 2001).

The discussion inevitably moved to the analogous problem regarding the
origin and constancy of the IMF.  While some observations, for example,
inferences of star formation rates in distant galaxies, are most
consistent with a top-heavy IMF, there is little direct evidence
locally for variations.  Resolved observations of stellar clusters and
unresolved observations of galaxies, for example, the existence of the
Fundamental Plane, almost all point to minimal variations in the IMF.

\section{The Fundamental Question}

In the above cycle $\rm (A)\rightarrow (B)\rightarrow (C)\rightarrow
(A)$, is one of these three issues more truly fundamental than the
other two?  For example, it is hard to conclude that Question (B), the
effect of feedback, is fundamental since feedback cannot occur without
stars.  On the other hand, without feedback-induced properties like
turbulence and metallicity, the ISM would be drastically different
than is found in present star-forming conditions.  Thus primordial
conditions may be excluded from this cycle in any case.  Furthermore,
preliminary evidence emerging from the above discussion shows that the
$N_*^{-2}$ cluster membership function and signature {\sc H\thinspace ii} LF are
remarkably constant; thus the short, radical(!) answer to Question (C)
might be that ISM properties {\it don't} affect and the {\sc H\thinspace ii} LF and
the cluster membership function.

For our present state of knowledge, Question (A) therefore emerges from
this Discussion as outstanding from the cycle.  We might reword it:
{\bf ``Is
the slope of the cluster membership function universal?''}  The extreme
conditions of the most luminous star-forming regions naively would seem to
offer a regime where variations might be found, analogous to the
suggestions that this energetic regime promotes an altered, top-heavy
stellar IMF.  To date, however, the evidence does not show any such
variations in the $N_*^{-2}$ law.  What is the origin of this law, and
does it have any relation to ISM properties?  We hope future discussions
and arguments will focus on, and resolve, this issue.

\acknowledgements

Our congratulations to the organizers for such a beautiful meeting. 
We suppose Roberto made it easier by having such a bunch of good friends...
We thank Elena Terlevich for taking notes during this
discussion, which have been very useful for us.  Thanks also to
Guillermo Tenorio-Tagle for reading the manuscript.

\end{document}